\documentclass[preprint,12pt]{elsarticle}

\usepackage{amssymb}
\usepackage{amsmath} 

\begin{document}

\begin{frontmatter}



\title{Ab Initio Prediction of Large Thermoelectric Effect in Distorted Heusler Alloy Ti-Fe-Sb Compound}


\author[inst1]{Rifky Syariati}

\affiliation[inst1]{organization={Nanomaterials Research Institute, Kanazawa University},
            addressline={Kakumamachi}, 
            city={Kanazawa},
            postcode={9201192}, 
            state={Ishikawa},
            country={Japan}}

\author[inst2]{Athorn Vora-ud}
\affiliation[inst2]{organization={Center of Excellence for Alternative Energy, Research and Development Institution,  Sakon Nakhon Rajabhat University},
            addressline={680 Nittayo Road}, 
            city={ Mueang District},
            postcode={47000}, 
            state={Sakon Nakhon},
            country={Thailand}}

\author[inst1]{Fumiyuki Ishii}

\author[inst2]{Tosawat Seetawan}

\begin{abstract}

The thermoelectric figure of merit of the Heusler alloy TiFe$_{1.5}$Sb were investigated by first-principles calculations of lattice thermal conductivity. The electronic thermal conductivity, electrical conductivity, and Seebeck coefficient are calculated by semi-classical Boltzmann transport theory. TiFe$_{1.5}$Sb was found to be thermally and dynamically stable, as confirmed by its phonon dispersion. Additionally, the small phonon band gap between acoustic and optical modes enhances phonon scattering, leading to a low lattice thermal conductivity of 0.703 W/mK at 300 K. Our study also reveals that TiFe$_{1.5}$Sb is a non-magnetic semiconductor. Notably, it demonstrates a significant longitudinal thermoelectric effect, with a Seebeck coefficient of 359.4 $\mu$V/K at 300 K. The combination of low lattice thermal conductivity and a high Seebeck coefficient results in a high thermoelectric figure of merit (ZT) of 0.88 and 0.91 at 300 K and 500 K, respectively. These findings highlight the considerable potential of TiFe$_{1.5}$Sb as a promising material for thermoelectric device applications.
\end{abstract}

\begin{graphicalabstract}
\includegraphics[width=1.0\textwidth]{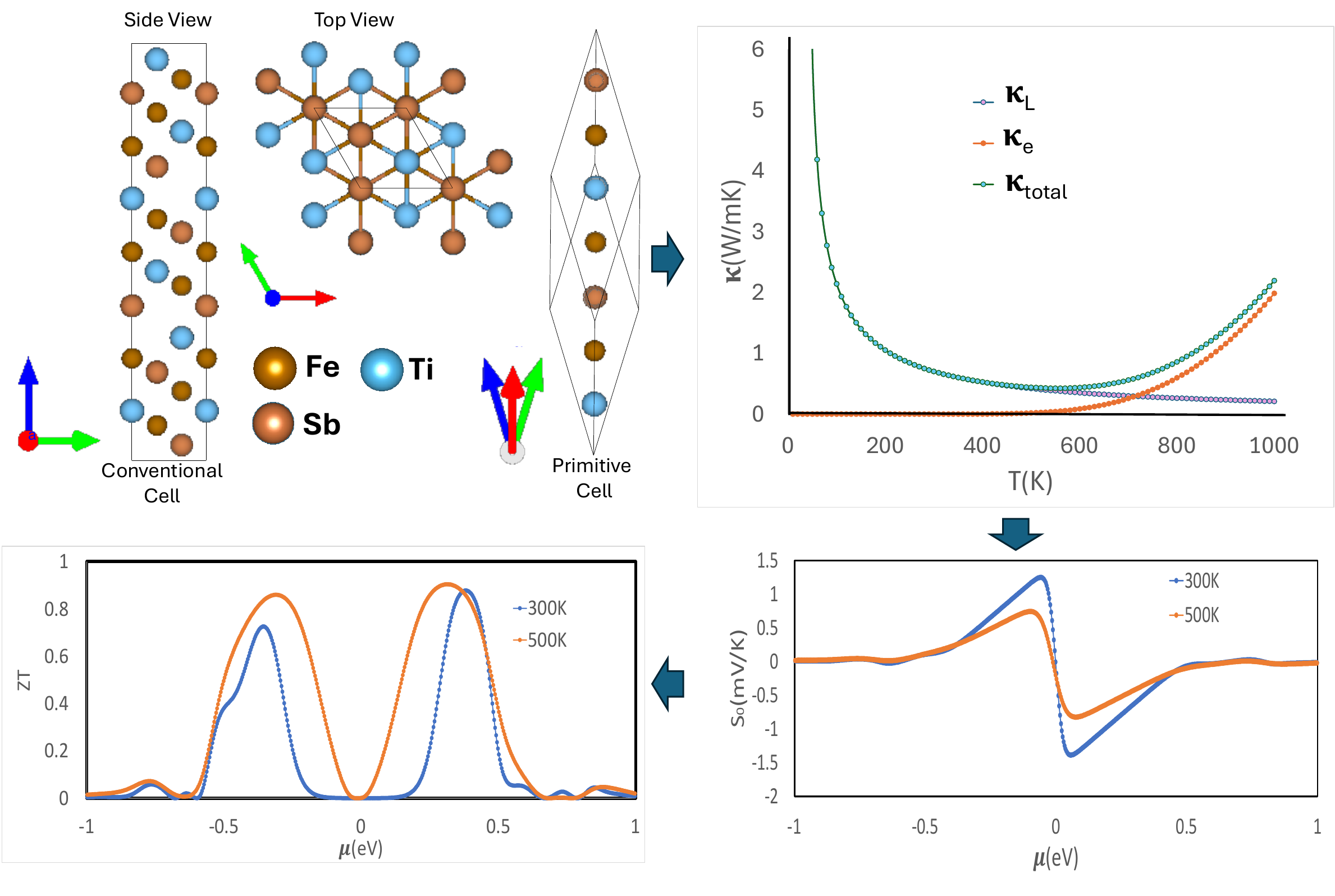}
\end{graphicalabstract}

\begin{highlights}
\item TiFe$_{1.5}$Sb has a low lattice thermal conductivity
\item Figure of merit of TiFe$_{1.5}$Sb is 0.88 at 300 K
\end{highlights}


\end{frontmatter}


\section{Introduction}
Currently, the world faces two significant energy-related challenges. The first is the energy crisis, prompting the exploration of alternative energy sources. The second concern is the environmental impacts associated with conventional energy consumption methods. These challenges are driving research into more efficient energy utilization strategies \cite{sharvini2018energy, hafez2023energy}. Thermoelectricity, which involves phenomena such as Seebeck and Nernst effects and lattice thermal conductivity, is emerging as a viable solution to these issues \cite{snyder2008complex,alam2013review,mizuguchi2019energy}.

The Seebeck effect, a foundational thermoelectric phenomenon, generates voltage across a material under a temperature gradient \cite{sun2015large, goldsmid2017seebeck}. When two different conductors or semiconductors are connected and subjected to varying temperatures at their junctions, an electromotive force (EMF) emerges \cite{goldsmid2017seebeck}. This EMF is produced as the electrons move from the warmer to the cooler region, creating a voltage difference \cite{lee2014seebeck}. To increase the Seebeck coefficient, strategies such as improving carrier mobility, increasing the effective mass of charge carriers, and fine-tuning carrier concentration are essential \cite{snyder2008complex, heremans2008enhancement}. These adjustments ensure sufficient electrical conductivity without compromising the Seebeck effect. In addition, the lattice thermal conductivity plays a crucial role in thermoelectricity efficiency. According to kinetic theory, lattice thermal conductivity is determined by the phonon's relaxation time, group velocity, and specific heat  \cite{toberer2011phonon}. Lower values of these parameters typically result in reduced lattice thermal conductivity. Phonons in a solid propagate independently under harmonic conditions, but anharmonic lattice vibrations lead to intrinsic phonon scattering, limiting their relaxation time \cite{heremans2015anharmonicity}. Materials with pronounced anharmonicity, such as SnSe \cite{zhao2014ultralow, li2015orbitally} and I-V-VI$_2$ compounds \cite{morelli2008intrinsically, ma2013glass}, exhibit a low $\kappa_L$. Furthermore, engineering microstructural defects is a common approach to minimize the phonon relaxation time and thus thermal conductivity, enhancing the material's thermoelectric efficiency \cite{mao2018self, zhu2021thermodynamic, chen2017lattice, qin2021substitutions, biswas2011strained}.


Transitioning from fundamental phenomena to practical applications, Heusler alloys have emerged as a prominent class of intermetallic compounds with distinctive crystal structures and magnetic properties, making them highly relevant in materials science, particularly for thermoelectric applications \cite{quinn2021advances, gui2023large, kumar2021thermoelectricity}. These alloys typically adopt an XYZ or X$_2$YZ composition, where X and Y are transition metals, and Z is a main group element \cite{felser2015heusler}. This structural arrangement allows for diverse electronic configurations, allowing a wide range of thermoelectric properties. Among these, the Ti-Fe-Sb system is of particular interest, as its structural variations exhibit distinct thermoelectric behaviors \cite{liu2019design, anand2022structural, kutorasinski2014application}. TiFe$_{1.5}$Sb, recognized as the dominant and stable phase within the Ti-Fe-Sb system, demonstrates significant thermoelectric potential due to its semiconducting attributes and the associated Seebeck effect \cite{naghibolashrafi2016synthesis}. Furthermore, the semiconducting properties of TiFe$_{1.5}$Sb are explained using the Slater–Pauling rule, which identifies nonmagnetic Heusler semiconductors with an average of six valence electrons per atom. This electron configuration induces a structural transition from cubic to hexagonal symmetry, which is linked to the material's low lattice thermal conductivity \cite{naghibolashrafi2016synthesis, wang2022discovery}.


In this study, we investigate the thermoelectric properties of the Ti-Fe-Sb alloy, with a particular focus on the composition TiFe$_{1.5}$Sb, which has been experimentally observed. Our results indicate that TiFe$_{1.5}$Sb exhibits semiconducting behavior, characterized by a high Seebeck coefficient. This remarkable Seebeck coefficient is primarily attributed to the presence of multiple electronic bands near the Fermi level, which increase the density of states and significantly affect carrier scattering mechanisms. Furthermore, TiFe$_{1.5}$Sb demonstrates low lattice thermal conductivity at room temperature, a critical factor for enhancing thermoelectric performance. The combination of a high Seebeck coefficient and low lattice thermal conductivity contributes to an improved figure of merit (ZT) for TiFe$_{1.5}$Sb. These findings suggest that TiFe$_{1.5}$Sb holds great promise for applications in advanced thermoelectric devices.

\section{Method}
We employed non-collinear density functional theory (DFT) calculations with the \textsc{OpenMX} ab-initio package to explore the structural and electronic characteristics of TiFe$_{1.5}$Sb ~\cite{openmx, LCPAO}. We applied the Perdew-Burke-Ernzerhof (PBE) version of the generalized gradient approximation (GGA) and norm-conserving pseudopotentials ~\cite{pseudopotential}. The wave functions were expanded using linear combinations of multiple pseudoatomic orbitals (PAOs) ~\cite{LCPAO, ozaki2004numerical, ozaki2005efficient}, with specific localized orbitals and radial cutoffs for Ti7.0-s3p2d1, Fe5.5H-s3p2d1, and Sb7.0-s3p2d2 with a radial cutoff of 7.0, 5.5, and 7.0 a.u. for the Ti, Fe, and Sb atom, respectively, validated by the delta gauge~\cite{lejaeghere2014error,lejaegherescience}. All structures were fully relaxed using the quasi-Newton algorithm~\cite{baker1986algorithm} until the residual force on each atom became smaller than $\sim0.5$ meV/\AA. We included spin-orbit coupling (SOC) in our fully relaxed non-collinear calculations ~\cite{theurich2001self}.

The lattice thermal conductivity ($\kappa_L$) is calculated using the Boltzmann transport equation with the relaxation time approximation, as shown below:

\begin{equation}\label{eq:7}
    \kappa_l^{\mu \nu}(T) = \frac{1}{\Omega N_q} \sum_{q,j} c_{qj}(T) v_{qj}^\mu v_{qj}^\nu \tau_{qj},
\end{equation}

where $\Omega$ is the volume of the primitive cell, $N_q$ is the number of wave vectors, $c_{qj}(T)$ represents the specific heat capacity of the phonon mode $qj$, $v_{qj}$ is the phonon group velocity, and $\tau_{qj}$ is the phonon relaxation time. In this study, a supercell size of $2 \times 2 \times 2$ (168 atoms) was used. To compute $\kappa_L$, it is essential to determine the interatomic force constants (IFCs), which describe interatomic interactions. 

The IFCs were optimized to match the atomic forces obtained from density functional theory (DFT) calculations. This optimization was performed using the least absolute shrinkage and selection operator (LASSO) technique, implemented in the \texttt{ALAMODE} code, to minimize overfitting \cite{tadano2015self}. The atomic forces were computed using the \texttt{OpenMX} code ~\cite{openmx, LCPAO}. Second-order harmonic terms and third-order anharmonic terms of the IFCs were calculated to account for phonon dispersion, group velocity, and specific heat, with the second-order IFCs generated from 35 displacement patterns of Heusler alloys.

To estimate the phonon relaxation time, we considered the dominant three-phonon scattering processes, which are significant above room temperature. Relaxation time was determined from the imaginary part of the self-energy phonon using third-order anharmonic IFCs ~\cite{yabuuchi2017first}. A total of 1108 displacement patterns generated by \texttt{ALAMODE} were used to compute the third-order IFCs, considering interactions up to the nearest neighbor with a cutoff radius of 8.1 Bohr. The convergence of the calculated $\kappa_L$ was verified using up to 1108 patterns with a cutoff radius of 8.1 Bohr, resulting in a negligible error of less than $10^{-3}$ W·K$^{-1}$·m$^{-1}$. 

Furthermore, to calculate the longitudinal thermoelectric coefficient of TiFe$_{1.5}$Sb, we use the semi-classical Boltzmann transport equation (BTE) under the relaxation time approximation (RTA) as implemented in BoltzTraP2 code \cite{madsen2018boltztrap2}. In the BTE, the Seebeck coefficient ($S_0$), the electrical conductivity ($\sigma$), and the thermal conductivity from the electrons contribution ($\kappa_e$) are defined as
\begin{equation}\label{eq:8}
S_0 = \frac{1}{eT} \frac{\mathcal{L}_1}{\mathcal{L}_0}, \quad \sigma = \mathcal{L}_0, \quad \kappa_e = \frac{1}{e^2T} \left[ \mathcal{L}_2 - \frac{\mathcal{L}_1^2}{\mathcal{L}_0} \right],
\end{equation} where e is the kernel $\mathcal{L}_{\alpha}$ of the generalized transport coefficient:

\begin{equation}\label{eq:9}
\mathcal{L}_\alpha = e^2 \sum_{nk} \tau_{nk}^2 (\epsilon_{nk} - \mu)^\alpha \left(-\frac{\partial f_{nk}}{\partial \epsilon_{nk}}\right), \quad \alpha = 0, 1, 2.
\end{equation} Besides that, the electron relaxation time ($\tau_{nk}$) is calculated using an empirical formula based on the carrier concentration in the function of chemical potential ($\mu$) which can be written as \cite{mili2022simple}
\begin{equation}\label{eq:10}
\tau_{nk} = \frac{qS_0h}{k_b^2T}
\end{equation} where \textit{q} is electron charge, \textit{h} is Planck constant, $k_B$ is Boltzmann constant, and T is temperature. By using all the output from \texttt{ALAMODE} and BoltzTraP2, the figure of merit (ZT) can be calculated as shown below
\begin{equation}\label{eq:11}
ZT = \frac{PF}{\kappa_{total}}T
\end{equation} where $PF=\sigma S^2$ is power factor and $\kappa_{total}=\kappa_L + \kappa_e$.

\section{Result and Discussion}
\begin{figure}
    \centering
    \includegraphics[width=0.65\textwidth]{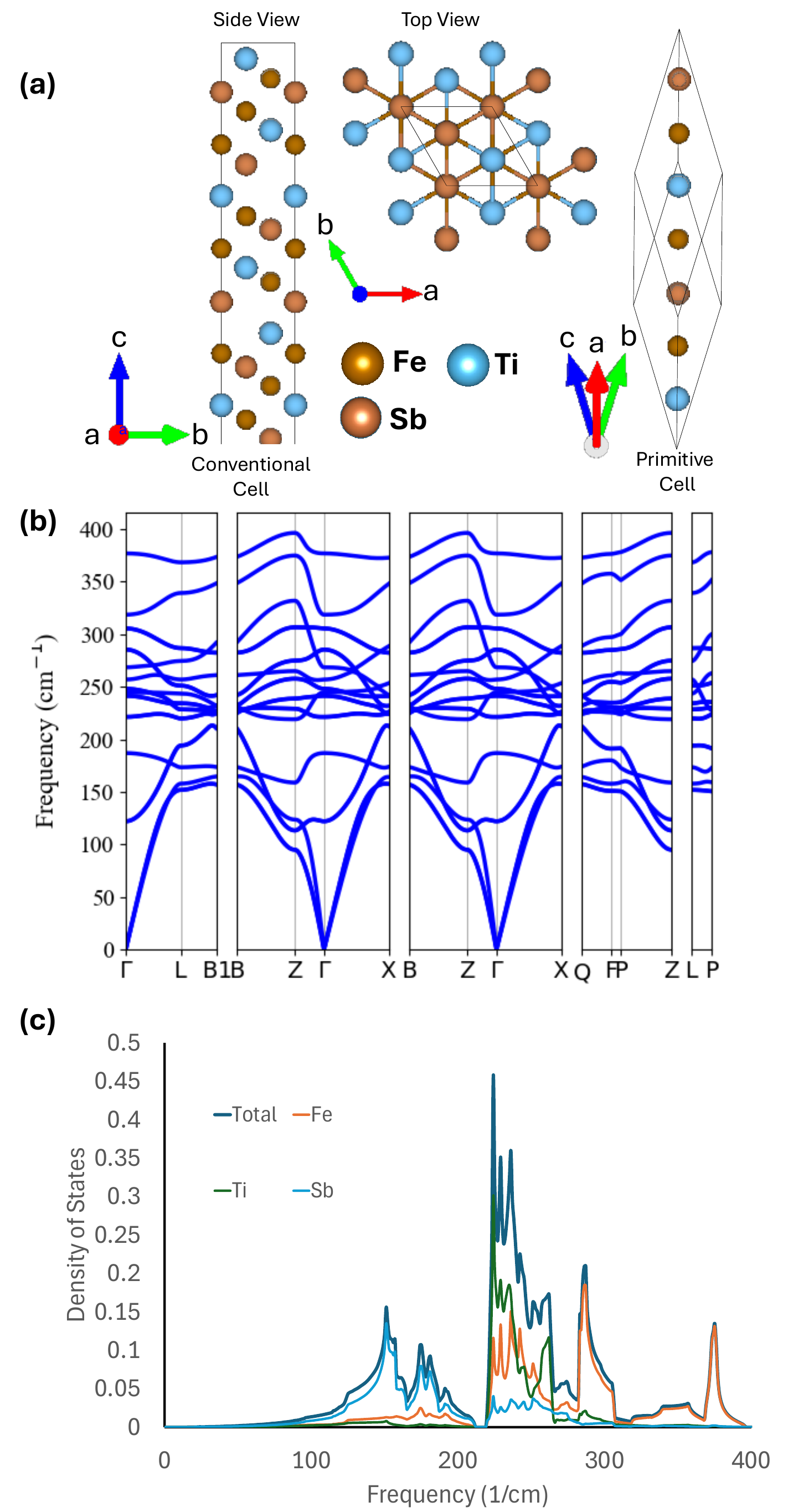}
    \caption{(a) Crystal structure, (b) phonon dispersion, and (c) the phonon density of states of TiFe$_{1.5}$Sb. Yellow, orange, red, and black lines indicate Fe, Sb, Ti, and total density of states, respectively. The crystal models are visualized by using VESTA~\cite{momma2011vesta}.}
    \label{fig:structure}
\end{figure}


Based on experimental results, TiFe$_{1.5}$Sb is the predominant phase in the Ti-Fe-Sb Heusler alloy compound \cite{naghibolashrafi2016synthesis}. The defective Heusler phase of TiFe$_{1.5}$Sb shows a significant structural shift, changing from a cubic to a hexagonal phase due to intrinsic defects. This structural change modifies the space group from \textit{F43m} to \textit{R3m}, as illustrated in Fig. \ref{fig:structure}(a). This transition is associated with stabilization of the TiFe$_{1.5}$Sb structure by preventing iron vacancies (NV = 0), which can be represented as Ti$^{4+}$Fe$^{2-}$Fe$_{0.5}^{2+}$Sb$^{3-}$ \cite{anand2022structural}. Additionally, TiFe$_{1.5}$Sb can be denoted as Ti$_2$Fe$_3$Sb$_2$, potentially forming from TiFeSb + TiFe$_2$Sb \cite{ti2023bonding}. Our calculations show that the lattice constant (\textit{a}) of TiFe$_{1.5}$Sb is 4.217 Å, which is consistent with previous theoretical and experimental findings \cite{naghibolashrafi2016synthesis, ti2023bonding}. TiFe$_{1.5}$Sb exhibits weak paramagnetic behavior due to its atomic structure. To further evaluate the thermal stability of this structure, its phonon dispersion needs to be examined.


The phonon dispersion of TiFe$_{1.5}$Sb, calculated from harmonic interatomic force constants (IFCs), is shown in Fig. \ref{fig:structure}(b). Since TiFe$_{1.5}$Sb contains three types of atoms, its phonon dispersion includes a total of 14 branches: four acoustic modes in the low-frequency region and ten optical modes in the high-frequency region. The absence of negative frequencies confirms the thermal stability of TiFe$_{1.5}$Sb. To understand the contribution of each atom type to the phonon dispersion, Fig. \ref{fig:structure}(c) provides the projected density of states (DOS). The low frequency range, from 100 to 200 cm$^{-1}$, mainly contributes to thermal transport and is dominated by vibrations of Sb atoms. This is due to the heavier mass of Sb compared to Ti and Fe, resulting in lower vibrational frequencies. In the higher-frequency optical modes, Fe and Ti atoms contribute more significantly: Ti atoms mainly from 220 to 275 cm$^{-1}$ and Fe atoms from 275 to 400 cm$^{-1}$. Furthermore, Fig. \ref{fig:structure}(b) and \ref{fig:structure}(c) also reveal a small band gap between the acoustic and optical modes. This small gap suggests that $\kappa_L$ may be low, as frequent phonon scattering can occur between these modes.

\begin{figure}
    \centering
    \includegraphics[width=1.0\textwidth]{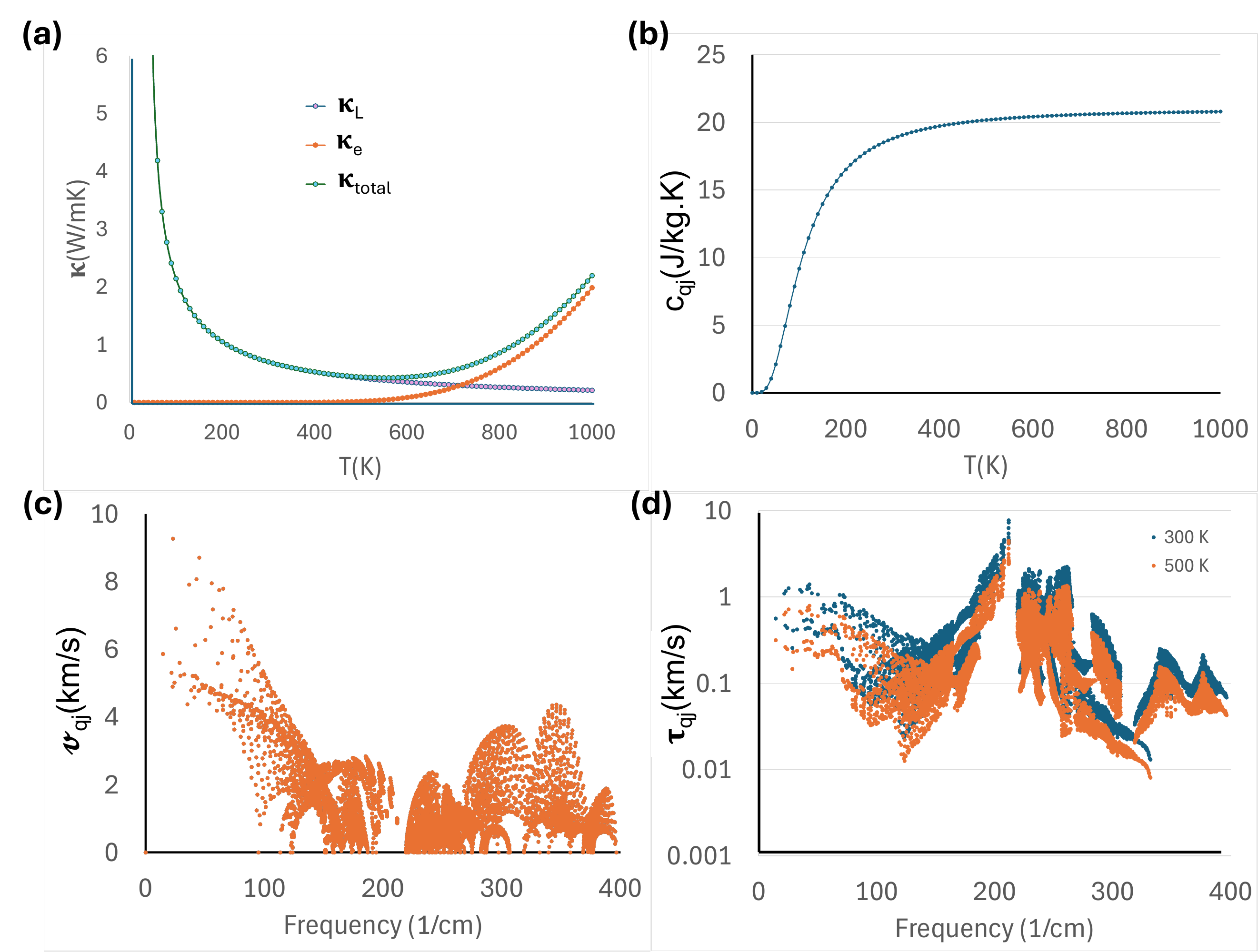}
    \caption{(a) Temperature dependence of thermal conductivity of TiFe$_{1.5}$Sb. Purple, orange, and blue solid point lines represent the phonon ($\kappa_L$), electron ($\kappa_e$), and total ($\kappa_{total}$) of thermal conductivity, respectively. (b) Specific heat capacity at constant volume as a function of temperature. (c) Phonon group velocity across the phonon spectrum, indicating the speed at which vibrational energy propagates. (d) Phonon relaxation time at 300 K and 500 K, illustrating the rate of energy dissipation for phonons at these temperatures.}
    \label{fig:lattice_thermal}
\end{figure}


The temperature dependence of $\kappa_L$, $\kappa_e$, and $\kappa_{total}$ is shown in Fig. \ref{fig:lattice_thermal}(a). The influence of the small band gap between acoustic and optical modes is evident here, leading to low $\kappa_L$. At 300 K and 500 K, $\kappa_L$ values are 0.703 and 0.422 W/mK, respectively. These values are comparable at room temperature to those found in materials such as SnSe, AgBiX$_2$ (X = S, Se, Te), Na$_2$KSb, K$_2$CsSb, LiAlTe$_2$, and LiGaTe$_2$ \cite{xiao2016origin, mandal2023physical, singh2022low}. Furthermore, Fig. \ref{fig:lattice_thermal}(a) illustrates that $\kappa_L$ decreases with increasing temperature. Besides that, Fig. \ref{fig:lattice_thermal}(a) also show that $\kappa_{total}$ mostly is generated from $\kappa_L$ than $\kappa_e$ at T $\leq$ 700 K.



$\kappa_L$ is related to $c_{qj}$, as shown in Fig. \ref{fig:lattice_thermal}(b). At low temperatures (near 0 K), $c_{qj}$ increases rapidly with temperature because only low-energy phonons (long-wavelength) are initially excited. This leads to a lower $c_{qj}$ that increases as more phonon modes are thermally activated. As the temperature continues to increase, $c_{qj}$ gradually approaches a plateau and becomes nearly constant, reflecting the full excitation of the available phonon modes in TiFe$_{1.5}$Sb. Above 500 K, $c_{qj}$ appears to saturate around 20 $J/(kg.K)$, suggesting that all phonon modes are fully populated and further temperature increases no longer significantly affect $c_{qj}$. This saturation indicates that the material has reached its maximum capacity for storing thermal energy in phonons.

$\nu_{qj}$ analysis for TiFe$_{1.5}$Sb as shown on Fig. \ref{fig:lattice_thermal}(c) shows that in the low frequency range (0 - 100 cm$^{-1}$), the $\nu_{qj}$ are relatively high, reaching up to approximately 8-9 km/s, indicating the presence of acoustic phonon modes, which are primary carriers of heat due to their ability to propagate energy efficiently. As frequency increases (100 - 200 cm$^{-1}$), the $\nu_{qj}$ begins to decrease, marking the transition to optical phonon modes, which have lower $\nu_{qj}$ and contribute less to $\kappa_L$ due to increased scattering. In the high-frequency region (200 - 400 cm$^{-1}$), the $\nu_{qj}$ drop further, typically staying below 2 $km/s$, consistent with the behavior of optical modes that have limited thermal transport capabilities. The small gap between the acoustic and optical modes suggests that some phonon-phonon scattering may occur, which would reduce $\kappa_L$. In general, the high $\nu_{qj}$ of the acoustic modes indicates their importance in thermal transport, while the low $\nu_{qj}$ and increased scattering in the optical modes help to keep the $\kappa_L$ of TiFe$_{1.5}$Sb relatively low.

The low $\kappa_L$ in TiFe$_{1.5}$Sb also can be explained by $\tau_{qj}$ as depicted in Fig. \ref{fig:lattice_thermal}(d). Fig. \ref{fig:lattice_thermal}(d) shows that, at both 300 K and 500 K, $\tau_{qj}$ are longer at low frequencies (up to 10 ps), where acoustic phonons dominate, allowing these phonons to travel longer distances with minimal scattering and contributing significantly to $\kappa_L$. As frequency increases, particularly in the mid to high frequency range (100 - 400 cm$^{-1}$), $\tau_{qj}$ decrease rapidly, falling below 1 ps, reflecting the greater scattering susceptibility of optical phonons, which contribute less to heat transport. At 500 K, $\tau_{qj}$ are generally shorter than at 300 K, indicating that phonon-phonon scattering increases with temperature, reducing $\kappa_L$. This temperature-dependent reduction in $\tau_{qj}$, combined with shorter lifetimes for high-frequency phonons, supports the tendency of TiFe$_{1.5}$Sb towards lower $\kappa_L$ at elevated temperatures.

\begin{figure}
    \centering
    \includegraphics[width=1.0\textwidth]{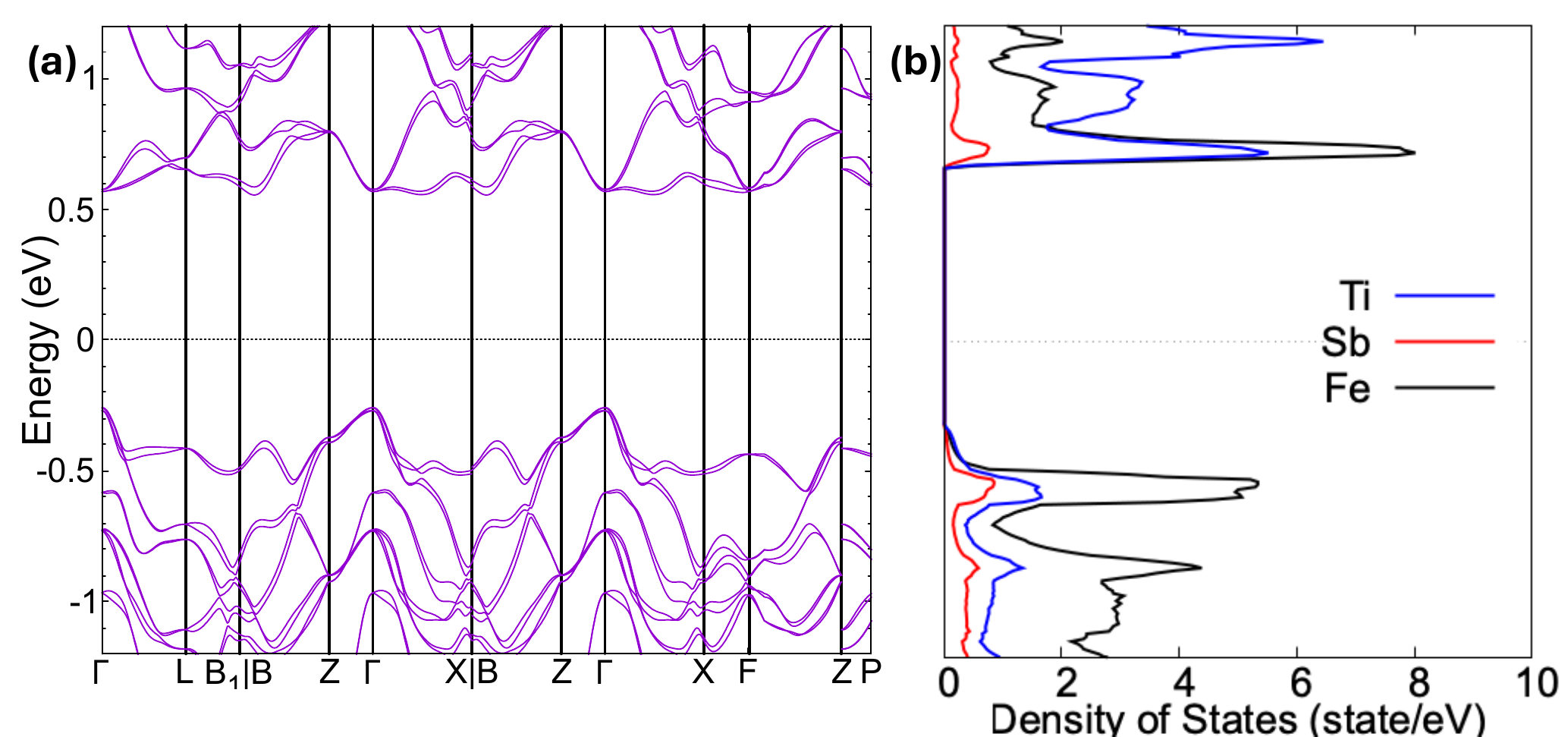}
    \caption{(a) Electronic band structure of the Heusler alloy TiFe$_{1.5}$Sb near the Fermi level, including the effects of spin-orbit coupling (SOC). (b) Projected density of states (PDOS) with SOC, showing the contribution of different atomic orbitals near the Fermi level, which is set to 0 eV.}
    \label{fig:band_structure}
\end{figure}

Since the $\kappa_L$ of TiFe$_{1.5}$Sb is small, there is potential for this material to achieve a high ZT. To evaluate ZT, it is essential to investigate the electronic structure of TiFe$_{1.5}$Sb. The electronic properties of Heusler alloys within the Ti-Fe-Sb system are visually depicted in Fig. \ref{fig:band_structure}. The electronic structure of TiFe$_{1.5}$Sb, displayed in Fig. \ref{fig:band_structure}(a), reveals semiconducting ground states with a band gap of approximately 0.83 eV. The origin of this band gap and the semiconductor properties of TiFe$_{1.5}$Sb are electron redistributions between the substructures of half-Heusler TiFeSb and full-Heusler TiFe$_2$Sb within the atomic formation of TiFe$_{1.5}$Sb. This band gap and semiconducting behavior result align with previous studies on TiFe$_{1.5}$Sb \cite{naghibolashrafi2016synthesis, ti2023bonding}. In contrast, due to its semiconducting behavior and weak paramagnetism, TiFe$_{1.5}$Sb might manifest a substantial longitudinal thermoelectric effect.

\begin{figure}
    \centering
    \includegraphics[scale=0.415]{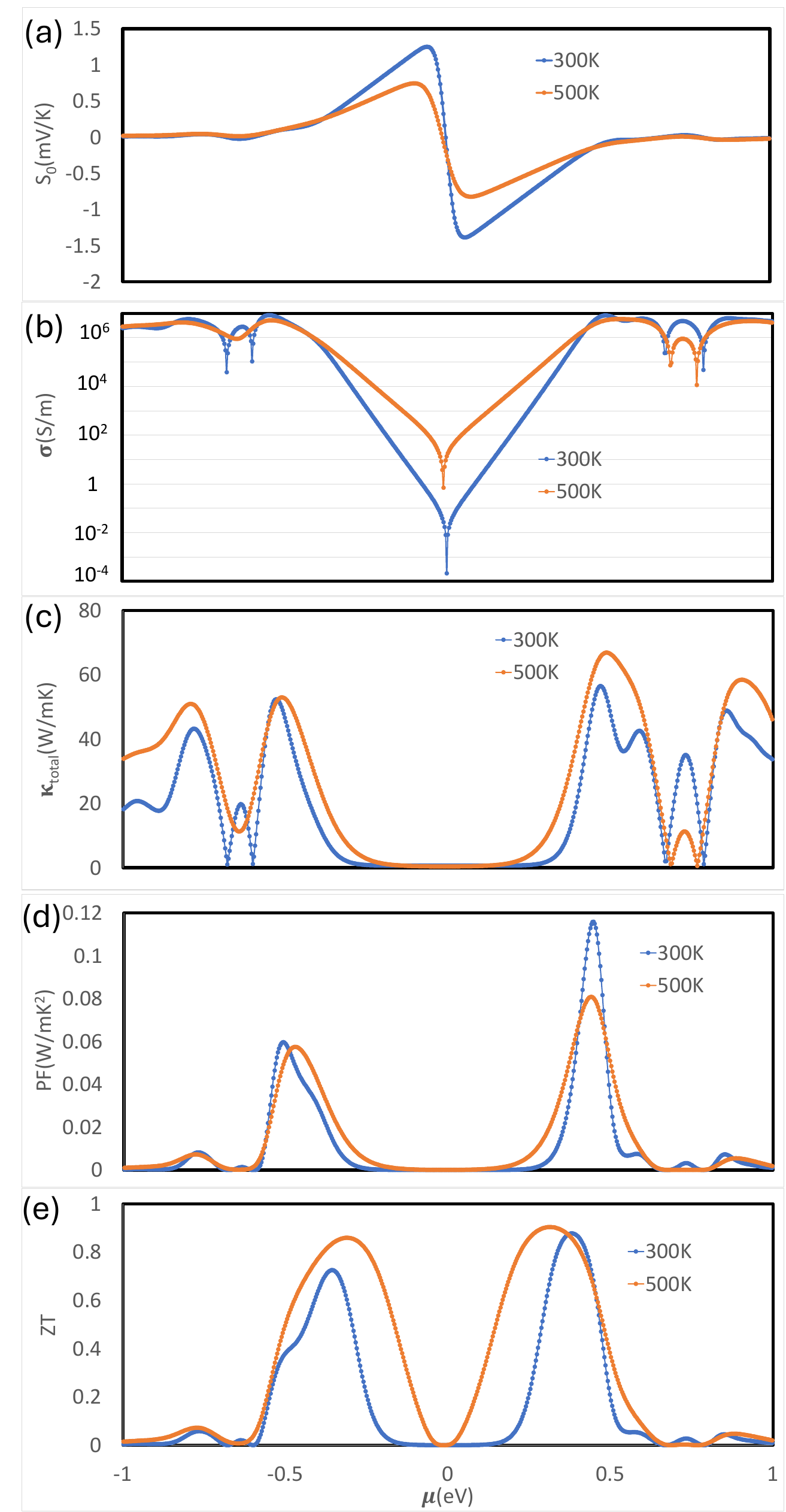}
    \caption{Chemical potential dependence of TE properties which consist of (a) Seebeck coefficient ($S_0$), (b) electrical conductivity ($\sigma$), (c) thermal conductivity total ($\kappa_{total}$), (d) power factor (PW), and (e) figure of merit (ZT) of Heusler alloys TiFe$_{1.5}$Sb at T=300K and T=500K, which are marked by blue and red lines, respectively.}
    \label{fig:seebeck}
\end{figure}

TiFe$_{1.5}$Sb exhibits significant longitudinal TE characteristics. This is illustrated in Fig. \ref{fig:seebeck}(a), highlighting the substantial S$_0$ in the carrier doping region. Specifically, at 300 K, S$_0$ reaches values as high as 1250 $\mu$V/K and 1387 $\mu$V/K for carrier doping levels of 2.31 $\times$ 10$^{17}$ electrons/cm$^3$ and 4.16 $\times$ 10$^{17}$ holes/cm$^3$, respectively. Upon increasing the temperature to 500 K, there is a shift in the peak of S$_0$ in the chemical potential, along with a reduction in its magnitude. Under these conditions, the S$_0$ values are observed to be 740 $\mu$V/K for electron doping and 824 $\mu$V/K for hole doping.

The large S$_0$ in TiFe$_{1.5}$Sb can be attributed to its electronic structure, as analyzed through its band structure and density of states (DOS), as depicted in \ref{fig:band_structure}(b). The band structure reveals multiple bands at both the valence band maximum and the conduction band minimum. These multiple degenerate energy bands lead to an increase in the DOS at these critical points, which in turn allows for different carrier scattering mechanisms. The different carrier scattering mechanisms significantly enhance S$_0$.

Due to the semiconducting properties of TiFe$_{1.5}$Sb, $\sigma$ at the Fermi level is notably low, as shown in Fig. \ref{fig:seebeck}(b). This low $\sigma$ spans from $\mu=-0.3$ eV to $\mu=0.3$ eV at 300 K, and from $\mu=-0.25$ eV to $\mu=0.25$ eV at 500 K. The limited magnitude of $\sigma$ originates from the material's band gap, which is illustrated in Fig. \ref{fig:band_structure}(a). Consequently, there is a critical need to balance S$_0$ with $\sigma$. For example, Fig. \ref{fig:seebeck}(a) displays a high S$_0$ around $\mu=-0.05$ eV at 300 K, while Fig. \ref{fig:seebeck}(b) shows a corresponding low $\sigma$ at the same chemical potential. This imbalance adversely affects PF of TiFe$_{1.5}$Sb. However, this issue can be addressed through charge doping, which, according to Fig. \ref{fig:seebeck}(d), significantly enhances PF around $\mu=0.45$ eV and $\mu=-0.45$ eV by incorporating both hole and electron doping. Furthermore, $\kappa$ remains low near the band gap but increases beyond $\mu=0.3$ eV, as shown in Fig. \ref{fig:seebeck}(c). Adjusting $\mu$, a high dimensionless thermoelectric ZT can be achieved, reaching nearly 1 at both 300 K and 500 K near the Fermi level, as evidenced in Fig. \ref{fig:seebeck}(e) and shown in Table. \ref{tab:ZT}.

Table. \ref{tab:ZT} summarizes the thermoelectric properties based on the highest ZT of a TiFe$_{1.5}$Sb at two temperatures, 300 K and 500 K, highlighting changes in electron doping levels ($n_i$), $S_0$, $\sigma$, $\kappa_{total}$, and ZT. At 300 K, $n_i$ is $3.56 \times 10^{20} \text{ cm}^{-3}$, with a $S_0$ of -359.42 $\mu $V/K, $\sigma$ of $2.99 \times 10^6 \text{S/m}$, and $\kappa_{total}$ of 13.24 W/mK, resulting in a ZT of 0.88. At 500 K, $n_i$ slightly increases to $3.54 \times 10^{20} \text{ cm}^{-3}$, while $S_0$ increases to -391.62 $\mu $V/K, $\sigma$ decreases to $1.38 \times 10^6 \text{ S/m}$, and $\kappa_{total}$ reduces to 11.71 W/mK, leading to a higher ZT of 0.91. These results indicate that the TiFe$_{1.5}$Sb's thermoelectric efficiency requires adjustment of all parameters to obtain a high ZT. In addition to that, its efficiency also improves at higher temperatures, suggesting its potential effectiveness as a thermoelectric generator at elevated temperatures.


\begin{table}[]
\centering
\begin{tabular}{cccccc}
\hline
T(K) & $n_i$(cm$^{-3}$)        & $S_0$($\mu V/K$) & $\sigma$(S/m)        & $\kappa_{total}$ (W/mK) & ZT   \\ \hline
300  & 3.56 $\times$ 10$^{20}$ & -359.42        & 2.99 $\times$ 10$^5$ & 13.24                   & 0.88 \\
500  & 3.54 $\times$ 10$^{20}$ & -391.62        & 1.38 $\times$ 10$^5$ & 11.71                   & 0.91 \\ \hline \hline
\end{tabular}
\caption{The calculated electron doping ($n_i$), Seebeck coefficient ($S_0$), electrical conductivity ($\sigma$), and total of thermal conductivity ($\kappa_{total}$) of TiFe$_{1.5}$Sb based on the highest ZT at T=300 K and T=500 K}
\label{tab:ZT}
\end{table}

\section{Conclusion}
TiFe$_{1.5}$Sb demonstrates a significant longitudinal thermoelectric effect, largely attributed to its low lattice thermal conductivity, which is measured at 0.703 W/mK at 300 K and decreases to 0.422 W/mK at 500 K. In addition, TiFe$_{1.5}$Sb exhibits an exceptionally high Seebeck coefficient ($S_0$), achieving values of -359.42 $\mu$V/K at charge doping levels of 3.56 × 10$^{20}$ electrons/cm$^3$ at 300 K. These properties enable TiFe$_{1.5}$Sb to achieve a high thermoelectric figure of merit (ZT) of approximately 0.88 at 300 K with electron doping. This combination of low lattice thermal conductivity and a high Seebeck coefficient highlights the potential of TiFe$_{1.5}$Sb for thermoelectric device applications.

\section{Acknowledgment}
This work was supported by e-ASIA JRP, Japan Science and Technology Agency (JST) Strategic International Collaborative Research Program (JST SICORP) with Grant Number JPMJSC21E3. The computation in this work was done using the facilities of the Supercomputer Center, the Institute for Solid State Physics, the University of Tokyo. The authors sincerely appreciate the discussions with Prof. G. Jeffrey Snyder on the thermal conductivity in TiFe$_{1.5}$Sb, which provided valuable insights that greatly contributed to this work.

 \bibliographystyle{elsarticle-num} 
 \bibliography{cas-refs}

\begin{thebibliography}{10}
\expandafter\ifx\csname url\endcsname\relax
  \def\url#1{\texttt{#1}}\fi
\expandafter\ifx\csname urlprefix\endcsname\relax\def\urlprefix{URL }\fi
\expandafter\ifx\csname href\endcsname\relax
  \def\href#1#2{#2} \def\path#1{#1}\fi

\bibitem{sharvini2018energy}
S.~R. Sharvini, Z.~Z. Noor, C.~S. Chong, L.~C. Stringer, R.~O. Yusuf, Energy consumption trends and their linkages with renewable energy policies in east and southeast asian countries: Challenges and opportunities, Sustainable Environment Research 28~(6) (2018) 257--266.

\bibitem{hafez2023energy}
F.~S. Hafez, B.~Sa'di, M.~Safa-Gamal, Y.~Taufiq-Yap, M.~Alrifaey, M.~Seyedmahmoudian, A.~Stojcevski, B.~Horan, S.~Mekhilef, Energy efficiency in sustainable buildings: a systematic review with taxonomy, challenges, motivations, methodological aspects, recommendations, and pathways for future research, Energy Strategy Reviews 45 (2023) 101013.

\bibitem{snyder2008complex}
G.~J. Snyder, E.~S. Toberer, Complex thermoelectric materials, Nature Materials 7~(2) (2008) 105--114.

\bibitem{alam2013review}
H.~Alam, S.~Ramakrishna, A review on the enhancement of figure of merit from bulk to nano-thermoelectric materials, Nano Energy 2~(2) (2013) 190--212.

\bibitem{mizuguchi2019energy}
M.~Mizuguchi, S.~Nakatsuji, \href{https://doi.org/10.1080/14686996.2019.1585143}{Energy-harvesting materials based on the anomalous {Nernst} effect}, Science and Technology of Advanced Materials 20~(1) (2019) 262--275.
\newblock \href {https://doi.org/10.1080/14686996.2019.1585143} {\path{doi:10.1080/14686996.2019.1585143}}.
\newline\urlprefix\url{https://doi.org/10.1080/14686996.2019.1585143}

\bibitem{sun2015large}
P.~Sun, B.~Wei, J.~Zhang, J.~M. Tomczak, A.~Strydom, M.~S{\o}ndergaard, B.~B. Iversen, F.~Steglich, Large seebeck effect by charge-mobility engineering, Nature Communications 6~(1) (2015) 7475.

\bibitem{goldsmid2017seebeck}
H.~J. Goldsmid, H.~Goldsmid, The {Seebeck} and {Peltier} effects, The Physics of Thermoelectric Energy Conversion (2017) 2053--2571.

\bibitem{lee2014seebeck}
E.-S. Lee, S.~Cho, H.-K. Lyeo, Y.-H. Kim, Seebeck effect at the atomic scale, Physical Review Letters 112~(13) (2014) 136601.

\bibitem{heremans2008enhancement}
J.~P. Heremans, V.~Jovovic, E.~S. Toberer, A.~Saramat, K.~Kurosaki, A.~Charoenphakdee, S.~Yamanaka, G.~J. Snyder, Enhancement of thermoelectric efficiency in {PbTe} by distortion of the electronic density of states, Science 321~(5888) (2008) 554--557.

\bibitem{toberer2011phonon}
E.~S. Toberer, A.~Zevalkink, G.~J. Snyder, Phonon engineering through crystal chemistry, Journal of Materials Chemistry 21~(40) (2011) 15843--15852.

\bibitem{heremans2015anharmonicity}
J.~P. Heremans, The anharmonicity blacksmith, Nature Physics 11~(12) (2015) 990--991.

\bibitem{zhao2014ultralow}
L.-D. Zhao, S.-H. Lo, Y.~Zhang, H.~Sun, G.~Tan, C.~Uher, C.~Wolverton, V.~P. Dravid, M.~G. Kanatzidis, Ultralow thermal conductivity and high thermoelectric figure of merit in {SnSe} crystals, Nature 508~(7496) (2014) 373--377.

\bibitem{li2015orbitally}
C.~W. Li, J.~Hong, A.~F. May, D.~Bansal, S.~Chi, T.~Hong, G.~Ehlers, O.~Delaire, Orbitally driven giant phonon anharmonicity in {SnSe}, Nature Physics 11~(12) (2015) 1063--1069.

\bibitem{morelli2008intrinsically}
D.~Morelli, V.~Jovovic, J.~Heremans, Intrinsically minimal thermal conductivity in cubic {I-V-VI$_2$} semiconductors, Physical Review Letters 101~(3) (2008) 035901.

\bibitem{ma2013glass}
J.~Ma, O.~Delaire, A.~May, C.~Carlton, M.~McGuire, L.~VanBebber, D.~Abernathy, G.~Ehlers, T.~Hong, A.~Huq, et~al., Glass-like phonon scattering from a spontaneous nanostructure in {AgSbTe$_2$}, Nature Nanotechnology 8~(6) (2013) 445--451.

\bibitem{mao2018self}
J.~Mao, J.~L. Niedziela, Y.~Wang, Y.~Xia, B.~Ge, Z.~Liu, J.~Zhou, Z.~Ren, W.~Liu, M.~K. Chan, et~al., Self-compensation induced vacancies for significant phonon scattering in {InSb}, Nano Energy 48 (2018) 189--196.

\bibitem{zhu2021thermodynamic}
Y.~Zhu, Z.~Han, F.~Jiang, E.~Dong, B.-P. Zhang, W.~Zhang, W.~Liu, Thermodynamic criterions of the thermoelectric performance enhancement in {Mg$_2$Sn} through the self-compensation vacancy, Materials Today Physics 16 (2021) 100327.

\bibitem{chen2017lattice}
Z.~Chen, Z.~Jian, W.~Li, Y.~Chang, B.~Ge, R.~Hanus, J.~Yang, Y.~Chen, M.~Huang, G.~J. Snyder, et~al., Lattice dislocations enhancing thermoelectric {PbTe} in addition to band convergence, Advanced Materials 29~(23) (2017) 1606768.

\bibitem{qin2021substitutions}
C.~Qin, L.~Cheng, Y.~Xiao, C.~Wen, B.~Ge, W.~Li, Y.~Pei, Substitutions and dislocations enabled extraordinary n-type thermoelectric {PbTe}, Materials Today Physics 17 (2021) 100355.

\bibitem{biswas2011strained}
K.~Biswas, J.~He, Q.~Zhang, G.~Wang, C.~Uher, V.~P. Dravid, M.~G. Kanatzidis, Strained endotaxial nanostructures with high thermoelectric figure of merit, Nature Chemistry 3~(2) (2011) 160--166.

\bibitem{quinn2021advances}
R.~J. Quinn, J.-W.~G. Bos, Advances in half-{Heusler} alloys for thermoelectric power generation, Materials Advances 2~(19) (2021) 6246--6266.

\bibitem{gui2023large}
Z.~Gui, G.~Wang, H.~Wang, Y.~Zhang, Y.~Li, X.~Wen, Y.~Li, K.~Peng, X.~Zhou, J.~Ying, et~al., Large improvement of thermoelectric performance by magnetism in {Co}-based full-{Heusler} alloys, Advanced Science 10~(28) (2023) 2303967.

\bibitem{kumar2021thermoelectricity}
R.~Kumar, R.~Singh, Thermoelectricity and Advanced Thermoelectric Materials, Woodhead Publishing, 2021.

\bibitem{felser2015heusler}
C.~Felser, A.~Hirohata, Heusler alloys, Vol. 222, Springer, 2015.

\bibitem{liu2019design}
Z.~Liu, S.~Guo, Y.~Wu, J.~Mao, Q.~Zhu, H.~Zhu, Y.~Pei, J.~Sui, Y.~Zhang, Z.~Ren, Design of high-performance disordered half-{Heusler} thermoelectric materials using 18-electron rule, Advanced Functional Materials 29~(44) (2019) 1905044.

\bibitem{anand2022structural}
S.~Anand, G.~J. Snyder, Structural understanding of the {Slater--Pauling} electron count in defective {Heusler} thermoelectric {TiFe$_1.5$Sb} as a valence balanced semiconductor, ACS Applied Electronic Materials 4~(7) (2022) 3392--3398.

\bibitem{kutorasinski2014application}
K.~Kutorasinski, J.~Tobola, S.~Kaprzyk, Application of {Boltzmann} transport theory to disordered thermoelectric materials: {Ti (Fe, Co, Ni)Sb} half-{Heusler} alloys, Physica Status Solidi (a) 211~(6) (2014) 1229--1234.

\bibitem{naghibolashrafi2016synthesis}
N.~Naghibolashrafi, S.~Keshavarz, V.~I. Hegde, A.~Gupta, W.~Butler, J.~Romero, K.~Munira, P.~LeClair, D.~Mazumdar, J.~Ma, et~al., Synthesis and characterization of {Fe}-{Ti}-{Sb} intermetallic compounds: Discovery of a new slater-pauling phase, Physical Review B 93~(10) (2016) 104424.

\bibitem{wang2022discovery}
L.~Wang, Z.~Dong, S.~Tan, J.~Zhang, W.~Zhang, J.~Luo, Discovery of a {Slater--Pauling} semiconductor {ZrRu$_{1.5}$Sb} with promising thermoelectric properties, Advanced Functional Materials 32~(25) (2022) 2200438.

\bibitem{openmx}
T.~Ozaki~et al, http://www.openmx-square.org/.

\bibitem{LCPAO}
T.~Ozaki, Variationally optimized atomic orbitals for large-scale electronic structures, Physical Review B 67 (2003) 155108.

\bibitem{pseudopotential}
N.~Troullier, J.~L. Martins, Efficient pseudopotentials for plane-wave calculations, Physical Review B 43 (1991) 1993--2006.

\bibitem{ozaki2004numerical}
T.~Ozaki, H.~Kino, Numerical atomic basis orbitals from {H} to {Kr}, Physical Review B 69~(19) (2004) 195113.

\bibitem{ozaki2005efficient}
T.~Ozaki, H.~Kino, Efficient projector expansion for the ab initio {LCAO} method, Physical Review B 72~(4) (2005) 045121.

\bibitem{lejaeghere2014error}
K.~Lejaeghere, V.~Van~Speybroeck, G.~Van~Oost, S.~Cottenier, Error estimates for solid-state density-functional theory predictions: an overview by means of the ground-state elemental crystals, Critical Reviews in Solid State and Materials Sciences 39~(1) (2014) 1--24.

\bibitem{lejaegherescience}
K.~Lejaeghere, G.~Bihlmayer, T.~Bj{\"o}rkman, P.~Blaha, S.~Bl{\"u}gel, V.~Blum, D.~Caliste, I.~Castelli, S.~Clark, A.~Dal~Corso, et~al., Reproducibility in density functional theory calculations of solids, Science 351~(6280) (2016) aad3000.
\newblock \href {https://doi.org/10.1126/science.aad3000} {\path{doi:10.1126/science.aad3000}}.

\bibitem{baker1986algorithm}
J.~Baker, An algorithm for the location of transition states, Journal of Computational Chemistry 7 (1986) 385--395.

\bibitem{theurich2001self}
G.~Theurich, N.~A. Hill, Self-consistent treatment of spin-orbit coupling in solids using relativistic fully separable ab initio pseudopotentials, Physical Review B 64 (2001) 073106.

\bibitem{tadano2015self}
T.~Tadano, S.~Tsuneyuki, Self-consistent phonon calculations of lattice dynamical properties in cubic {SrTiO$_3$} with first-principles anharmonic force constants, Physical Review B 92~(5) (2015) 054301.

\bibitem{yabuuchi2017first}
S.~Yabuuchi, Y.~Kurosaki, A.~Nishide, N.~Fukatani, J.~Hayakawa, First-principles study on thermoelectric transport properties of {Ca$_3$Si$_4$}, Physical Review Materials 1~(4) (2017) 045405.

\bibitem{madsen2018boltztrap2}
G.~K. Madsen, J.~Carrete, M.~J. Verstraete, {BoltzTraP2}, a program for interpolating band structures and calculating semi-classical transport coefficients, Computer Physics Communications 231 (2018) 140--145.

\bibitem{mili2022simple}
I.~Mili, H.~Latelli, Z.~Charifi, H.~Baaziz, T.~Ghellab, A simple formula for calculating the carrier relaxation time, Computational Materials Science 213 (2022) 111678.

\bibitem{momma2011vesta}
K.~Momma, F.~Izumi, {VESTA3} for three-dimensional visualization of crystal, volumetric and morphology data, Journal of Applied Crystallography 44~(6) (2011) 1272--1276.

\bibitem{ti2023bonding}
Z.~Ti, J.~Zhu, S.~Guo, J.~Li, X.~Liu, Y.~Zhang, Bonding properties of rubik's-cube-like {Slater-Pauling Heusler} semiconductors for thermoelectrics, Physical Review B 108~(19) (2023) 195203.

\bibitem{xiao2016origin}
Y.~Xiao, C.~Chang, Y.~Pei, D.~Wu, K.~Peng, X.~Zhou, S.~Gong, J.~He, Y.~Zhang, Z.~Zeng, et~al., Origin of low thermal conductivity in {SnSe}, Physical Review B 94~(12) (2016) 125203.

\bibitem{mandal2023physical}
S.~Mandal, P.~Sarkar, Physical insights into the ultralow lattice thermal conductivity and high thermoelectric performance of bulk {LiMTe$_2$(M= Al, Ga)}, Journal of Materials Chemistry C 11~(40) (2023) 13691--13706.

\bibitem{singh2022low}
U.~Singh, S.~Singh, M.~Zeeshan, J.~van~den Brink, H.~C. Kandpal, Low lattice thermal conductivity in alkali metal based {Heusler} alloys, Physical Review Materials 6~(12) (2022) 125401.

\end{thebibliography}





\end{document}